\documentclass[useAMS,usenatbib,usegraphicx]{mn2e}
\usepackage[usenames]{color}
\newcommand{\arcdeg}{\degr}
\newcommand{\uv}{\mbox{$u$-$v$}}

\newcommand{\kms}{\mbox{km s$^{-1}$}}
\newcommand{\Jb}{\mbox{Jy bm$^{-1}$}}
\newcommand{\muJb}{\mbox{$\mu$Jy bm$^{-1}$}}
\newcommand{\Ra}[4]{\mbox{${#1}^{\rm h} \; {#2}^{\rm m} \; {#3}\fs{#4} $}}
\newcommand{\dec}[4]{\mbox{${#1}\arcdeg \; {#2}\arcmin \; {#3}\farcs{#4} $}}
\newcommand{\thobs}{\mbox{$\theta_{\rm obs}$}}
\newcommand{\Ekiso}{\mbox{$E_{\rm k,iso}$}}
\newcommand{\ergsHz}{\mbox{erg~s$^{-1}$~Hz$^{-1}$}}
\newcommand{\LXband}{\mbox{$L_{\rm 8.4\,GHz}$}}
\newcommand{\lesssim}{\mbox{\raisebox{-0.3em}{$\stackrel{\textstyle <}{\sim}$}}}
\newcommand{\gtrsim}{\mbox{\raisebox{-0.3em}{$\stackrel{\textstyle >}{\sim}$}}}

\newcommand{\tablenotemark}[1]{$^{\rm #1}$}
\newcommand{\tablenotetext}[2]{\noindent$^{\rm #1}$ #2\\}
\newcommand{\phn}{\phantom{1}}

\newcommand{\objectname}[1]{#1}
\newcommand{\slugcomment}[1]{\date{#1}}

\title{Radio limits on off-axis GRB afterglows and VLBI observations of SN 2003\lowercase{gk}}

\author[Bietenholz et al.]{M. F. Bietenholz$^{1,2}$,
F. De Colle$^3$, J. Granot$^4$,
N. Bartel$^2$ and A. M. Soderberg$^5$ \\
$^1$Hartebeesthoek Radio Observatory, PO Box 443, Krugersdorp, 
1740, South Africa\\
$^2$Dept.\ of Physics and Astronomy, York University, Toronto,
M3J~1P3, Ontario, Canada\\
$^3$Instituto de Ciencias Nucleares, Universidad Nacional Aut\'onoma
de M\'exico, A. P. 70-543 04510 D. F., Mexico\\
$^4$Department of Natural Sciences, The Open University of Israel, 1
University Road, P.O. Box 808, Ra'anana 43537, Israel\\
$^5$Harvard-Smithsonian Center for Astrophysics, 60
Garden Street, Cambridge, MA 02138, US
}

\begin{document}

\slugcomment{\today;  Accepted to MNRAS}

\pagerange{\pageref{firstpage}--\pageref{lastpage}} \pubyear{2011}
 
\maketitle

\label{firstpage}

\begin{abstract}
We report on a VLA survey for late-time radio emission
from 59 supernovae (SNe) of Type I b/c, which have been associated
with long-duration gamma-ray bursts (GRBs).  An ``off-axis'' GRB burst
(i.e. whose relativistic jet points away from us) is expected to have
late-time radio emission even in the absence of significant prompt
gamma-ray emission.  From our sample, we detected only SN~2003gk with
an 8.4-GHz flux density of $2260 \pm 130 \,\mu$Jy.  Our subsequent
VLBI observations of SN~2003gk, at an age of $\sim$8~yr, allowed us to
determine its radius to be $(2.4 \pm 0.4) \times 10^{17}$~cm, or $94
\pm 15$ light days.  This radius rules out relativistic expansion as
expected for an off-axis GRB jet, and instead suggests an expansion
speed of $\sim 10\:000$~\kms\ typical for non-relativistic
core-collapse supernovae. We attribute the late-onset radio emission
to interaction of the ejecta with a dense shell caused by episodic
mass-loss from the progenitor.

In addition, we present new calculations for the expected radio
lightcurves from GRB jets at various angles to the line of sight, and
compare these to our observed limits on the flux densities of the
remainder of our SN sample.  From this comparison we can say that only
a fraction of broadlined Type I b/c SNe have a radio-bright jet
similar to those seen for GRB afterglows at cosmological distances.
However, we also find that for a reasonable range of parameters, as
might be representative of the actual population of GRB events rather
than the detected bright ones, the radio emission from the GRB jets
can be quite faint, and that at present, radio observations do not
place strong constraints on off-axis GRB jets.
\end{abstract}


\section{Introduction}

Long-duration gamma-ray bursts (GRBs) are thought to involve the
highly directed relativistic ejection of material from a collapsing
massive star, in other words the formation of relativistic jets.  The
observed gamma-ray emission is produced within the outflow, outside of
the progenitor star but before there is significant deceleration by
the surrounding medium.  The ejected material interacts with the
circumstellar material (CSM) and strong shocks are produced.  These
shocks amplify the magnetic field and accelerate particles to
relativistic energies, the combination of which results in synchrotron
emission.
For reviews of GRBs see for example \citet{Piran2004},
\citet{Meszaros2006}, \citet{Granot2007} and \citet{GehrelsRF2009}.

A GRB is observed when such a jet is directed close to the line of
sight, and strong Doppler boosting is responsible for the
characteristic bright gamma-ray emission.  It follows from this model,
however, that the majority of GRB events have jets which are not
oriented near the line of sight, and therefore go undetected
\citep{Rhoads1997, Granot+2002, NakarPG2002}, with \citet{Frail+2001}
estimating that $>99$\% of GRB events go undetected.

However, on the basis of several nearby ($z < 0.3$) long-duration GRBs
it has been established that they are associated with supernovae (SNe)
of Type Ib/c, which are ones from a progenitor star that has lost much
of its envelope prior to the explosion \citep[e.g.][]{Galama+1998,
  Stanek+2003, Malesani+2004, Pian+2006, Cobb+2010, Starling+2011,
  Xu+2013}.  In addition, there is indirect evidence for this
association, such as large ongoing specific star formation rates at
the location of the GRBs in their host galaxies, or late-time red
bumps in their afterglow lightcurves \citep[see, e.g.][and references
  therein]{WoosleyB2006}.  These events therefore feature both highly
collimated relativistic jets giving rise to the GRB and an associated
more isotropic and non-relativistic SN explosion.  This duality
underlies the popular ``collapsar'' model \citep{Woosley1993,
  MacFadyenWH2001} in which a central engine (accreting, rapidly
spinning compact object) drives the relativistic jets while the
spherical SN explosion is powered by neutrinos.

GRB events, in addition to the gamma-ray emission, also produce
longer-lived emission at lower frequencies, called the afterglow.  In
particular, afterglows are often detected in radio \citep[see,
  e.g.,][]{vParadijsKW2000, Zhang2007}. As the GRB decelerates to
mildly or sub-relativistic speeds, it produces strong, nearly
isotropized, radio emission.  The radio emission is much less strongly
beamed than the gamma-ray emission, and so in the radio ``off-axis''
events are not significantly more difficult to detect than on-axis
ones at sufficiently late times.  Indeed, it has been shown that the
radio is probably the best wavelength range for detecting such
off-axis events \citep[e.g.][]{Paczynski2001, GranotL2003}.

Models of off-axis GRB events developed so far have typically showed
that, for angles to the line of sight of 30\arcdeg\ to 90\arcdeg, the
radio brightness peaks $\sim$1 to $\sim$2 years after the explosion
\citep{vEertenZM2010, GranotL2003}.  Future large-area surveys with
instruments such as ASKAP, LOFAR, MeerKAT, and of course SKA, will
likely detect such off-axis events in blind surveys. At present,
however, detection in a blind survey is challenging, and radio
observations are mostly restricted to follow-up observations of events
previously detected in other wavelengths \citep[see][and references
  therein]{ChandraF2012}.

Type I b/c SNe provide the ideal locations to search for off-axis GRB
events.  Since non-relativistic SNe can also produce radio emission,
some means of distinguishing the radio emission from putative off-axis
jet from that of the normal SN is needed.  In the models available at
present, the timescales of the two processes are notably different and
could therefore provide the necessary discriminant.  In particular,
the models suggested that radio emission from an off-axis event had a
relatively long interval between the explosion and the time of maximum
radio emission, so that for angles to the line of sight of
30\arcdeg\ to 90\arcdeg, the peak radio brightness typically occurs
$\sim$1 to $\sim2$ years after the explosion \citep{vEertenZM2010,
  GranotL2003}.
The radio emission from normal Type I b/c SNe, on the other
hand, has short timescales, with rise-times at 8.4~GHz of typically a
few weeks to months after the explosion, followed by a relatively
rapid decay with flux density, $S$ approximately $\propto t^{-1.5}$
\citep{ChevalierF2006, Weiler+2002}.
Off-axis GRB events are also distinguished by having high peak
spectral luminosities, generally $>10^{28}$~\ergsHz\ (8.4 GHz), while
those of normal Type I b/c SNe are mostly $<10^{28}$~\ergsHz, although
some very radio bright normal SNe are known \citep[see,
e.g.][]{Soderberg2007}, such as SN~2003L \citep{Soderberg+2005}.

The models therefore suggested that late-onset and luminous radio
emission from a Type I b/c SN could be used as a signpost of an
off-axis GRB event.  \citet{Soderberg+2006b} carried out a search for
such events, and observed 68 Type I b/c SNe in the radio at late
times.  They did not detect any radio emission and concluded that only
a fraction of $<10$\% of all Type I b/c SN are associated with
a bright GRB jet regardless of orientation.  They could also rule out,
at the 84\% confidence level, the hypothesis that all of the subset of
Type I b/c SNe which had broad absorption lines and are therefore
classified as ``broadlined'' (sometimes also termed ``hypernovae'')
such as SN~2003jd or SN~2002ap, are associated with GRB events.

Should late-onset radio emission be detected from a Type I b/c, and
important diagnostic would be the size of the radio emitting region.
If it is indeed due to a relativistic jet, sizes on the order of one
light-year are expected.  Non-relativistic SNe, however, expand
much more slowly, typically with initial speeds of $0.1\,c$ and
average speeds on the order of $10\:000$~\kms\ over periods of
$>1$~yr, and would therefore be an order of magnitude smaller.

Indeed there have been several Type I b/c SNe suspected of possibly
harbouring an off-axis GRB where subsequent VLBI observations showed
that there was {\em no} relativistic expansion \citep[see][for a
  recent review of VLBI observations of Type I b/c
  SNe]{SNIbc-VLBI}. SN~2001em showed late-onset radio emission
\citep{GranotR2004, SN2001em-1, SN2001em-2, Schinzel+2009}.  The late
turn-on radio emission from SN~2001em was subsequently interpreted as
radio emission produced by the interaction of normal
(non-relativistic) Type I~b/c ejecta with a massive and dense
circumstellar shell located at some distance from the progenitor,
produced by mass-loss from the latter, perhaps due do an eruptive
event like those of luminous blue variables
\citep{ChugaiC2006,Chevalier2007}. 
SN~2007bg and the supernova PTF 11gcj also showed late-onset
radio-emission \citep{Salas+2013, Corsi+2014} that is also attributed
to the interaction of a non-relativistically expanding shock with
dense shells of circumstellar material produced by episodic mass loss
of the progenitor.
In another case, SN~2007gr, it was initially claimed that the VLBI
observations implied relativistic expansion \citep{SN2007gr-Nature},
however, \citet{SN2007gr-Soderberg} subsequently suggested that all
the observations could be explained by an ordinary,
non-relativistically expanding SN.

In a third nearby Type I b/c SN, SN~2009bb, the high radio
luminosity also suggested relativistic ejection
\citep{SN2009bb_Nature}.  In this case the VLBI observations were
consistent with, but did not demand, mildly relativistic expansion
\citep{SN2009bb-VLBI}.

The radio lightcurve of a non-relativistic SN generally has a
steep rise, as the radio emission is initially absorbed by either
synchrotron self-absorption or, less commonly for Type I b/c SNe, by
free-free absorption due to an optically thick CSM \citep[see,
e.g.][]{Chevalier1998}.  Once the SN has become optically thin
at the frequency of interest, the lightcurve generally decays.  The
canonical model of \citet{Chevalier1982b}, which assumes power-law
radial density profiles for both the CSM and the ejecta, produces a
power-law decay in the lightcurve after the peak. The interval between
the explosion and the peak radio brightness is a function of the
observing frequency, generally being longer at lower frequencies.  It
should however, be noted, that significant departures from a strictly
power-law decline in the radio emission are relatively common (e.g.\
SN~2009bb \citealt{SN2009bb-VLBI}, SN~1996cr \citealt{Meunier+2013},
SN~1986J, \citealt{SN86J-1, SN86J-2}, SN~1979C \citealt{SN79C-Shell}).

As there are as yet no confirmed examples of off-axis GRB jets without
a detected gamma-ray signature
it would be very important to detect one, or even in the absence of a
detection to set limits on the fraction of Type I b/c SNe which are
associated with a relativistic ejection.  Earlier work by
\citet{Soderberg+2006b, SoderbergFW2004} and \citet{Gal-Yam+2006}
showed that at most a small fraction of Type I b/c can be associated
with bright jets typical of detected cosmological GRBs.  Nonetheless,
if the current paradigm of GRBs involving highly directed ejection is
correct, then for every observed GRB there must be many as yet
unobserved off-axis events.  Searching for late-time radio emission
seems a relatively promising way to detect such an event, despite the
knowledge that many Type I b/c SNe will have to be searched to obtain
one detection.

We therefore undertook a radio survey of Type I b/c SNe with ages
between 1 and 8 yrs to look for late-time radio emission using the
NRAO Very Large Array (VLA).  In Section \ref{svlasurvey} we describe
this survey and give the results. One object, SN~2003gk, was detected,
and in Section \ref{s2003gk} give discuss our followup VLA and VLBI
observations of SN~2003gk.  In Section \ref{smodels} we calculate new
modelled radio lightcurves for off-axis GRB jets, and compare them to
our observations in Section \ref{scompare}. We discuss the
implications of our results in Section \ref{sdiscuss}, and summarize
our conclusions in Section \ref{ssummary}.

\section{Survey for late-time radio emission from Type I b/c supernovae}
\label{svlasurvey}

\subsection{VLA survey observations}
\label{ssurveyobs}

We use the VLA to survey a sample of 59 Type I b/c SNe with
declinations $> -30\arcdeg$ and with ages between 1 and 8 yrs, for
late-time radio emission.  We chose ordinary Type I b/c SNe only at
distances $< 80$~Mpc, but we also observed several Type I b/c SNe of
the ``broadlined'' subtype which is most reliably associated with
GRBs out to slightly larger distances up to 120~Mpc.  Our sample is
not intended to be complete, merely representative.

We observed in two sessions of 4 hours each, on 2009 May 28 and May 29
(observing code AB1327).
The array was in the CnB transitional configuration, and we observed
with a total bandwidth of 100~MHz around a central frequency of
8.435~GHz.  
The data reduction was carried out in the standard way, using 3C~48
and 3C~286 as flux density calibrators on the two days respectively
(using the VLA 1999.2 flux density scale).  
Each SN was observed for $\sim$7.3 min, with phase calibration
derived from bracketing scans of a nearby compact calibrator sources.

After calibration, we imaged the SNe.  If there was sufficient
flux density in the field, self-calibration in phase was attempted.
However, the improvements in the SN images achieved by
self-calibration ranged from non-existent to insignificant, suggesting
that our initial phase-calibration is generally adequate for our
purposes.  Both the effective resolution and image background rms
values varied from one SN to the other.  The FWHM areas of the
convolving beam ranged from $\sim$3.5 to $\sim$12 square arcseconds,
and image rms values from $\sim$50 to $\sim$100 \muJb.  We can
consider the SN positions accurately known for our purposes,
since the coordinates of the SNe were obtained from optical
observations, and are usually accurate to an arcsecond or better, and
the position errors due to the VLA phase referencing are also expected
to by $<1$\arcsec, whereas the FWHM resolution of the radio
observations was mostly $>2 \arcsec$.

\subsection{VLA survey results}
\label{ssurveyresults}

We present our flux density measurements, or upper limits, for each of
the SNe in Table~\ref{tsne}.  As an uncertainty in the flux
density we take the background rms or the radio image.  With the
exception of SN~2003gk, which is discussed below, we detected none our
sample SNe.  Since, as mentioned, the SN positions are
accurately known, we take the brightness of the radio image at the
SN position as an estimate of the SN's flux density.  If
this flux density is less than the flux-density uncertainty, we give
only $3\sigma$ upper limit on the flux density in Table~\ref{tsne}.
In the cases where the flux density exceeds the image rms, we add the
estimate of the flux density as well as its uncertainty in addition to
the $3\sigma$ limit.

The only reliably detected source was SN~2003gk.  The observed
morphology is consistent with being unresolved (as expected), with a
flux density of $2260 \pm 130$~\muJb, where the uncertainty consists
of the image rms and a 5\% calibration uncertainty added in
quadrature. 

For the other 58 SNe, the $3\sigma$ upper limits given in
Table~\ref{tsne} are conservative upper limit on the flux density of
the SN since the presence of extended emission due to the galaxy
cannot be ruled out.  We note that in several cases, galactic emission
is clearly seen at the SN location, and our limit on the SN emission
is the limit on any compact emission in excess of the galactic
emission at the SN location.
No unresolved sources (except for SN~2003gk) were seen within several
arcseconds of the nominal locations of any of our SNe.

\begin{table*}
\scriptsize
\begin{minipage}[t]{\textwidth}
\caption{Observed Supernovae}
\label{tsne}
\begin{tabular}{l l l c c l l}
\hline
Supernova & Galaxy & Type\tablenotemark{a} &
$D$\tablenotemark{c}  & Age\tablenotemark{b} & 8.5-GHz Flux Density\tablenotemark{d} 
\\
    &      &     & {(Mpc)} & (yr) & ($\mu$Jy) &
\\
\hline
SN 2001ej & UGC 3829    & Ib        & 57 & 7.7 & $<145$ \\ 
SN 2001is & NGC 1961    & Ib        & 56 & 7.4 & $<161$ \\ 
SN 2002J  & NGC 3464    & Ic        & 56 & 7.4 & $<246 \phn (87 \pm 53)$ \\ 
SN 2002bl & UGC 5499    & Ic/BL     & 71 & 7.3 & $<357$\\ 
SN 2002cp & NGC 3074    & Ibc       & 76 & 7.2 & $<263$ \\ 
SN 2002hf & MCG-05-3-20 & Ic        & 76 & 6.6 & $<241$ \\  
SN 2002hn & NGC 2532    & Ic        & 75 & 6.6 & $<\phn93$\\ 
SN 2002ho & NGC 4210    & Ic        & 43 & 6.7 & $<266$ \\ 
SN 2002hy & NGC 3464    & Ib pec    & 56 & 6.6 & $<247$ \\ 
SN 2002hz & UGC 12044   & Ic        & 76 & 6.7 & $<195$ \\ 
SN 2002ji & NGC 3655    & Ib/c      & 28 & 6.5 & $<272$ \\ 
SN 2002jj & IC 340      & Ic        & 55 & 6.5 & $<208$ \\ 
SN 2002jp & NGC 3313    & Ic        & 55 & 6.6 & $<227$ \\ 
SN 2003H  & NGC 2207    & Ib pec    & 38 & 6.4 & $<188$ \\ 
SN 2003dr & NGC 5714    & Ib/c pec  & 38 & 6.2 & $<177$ \\ 
SN 2003gf & MCG-04-52-26& Ic        & 37 & 6.1 & $<213$ \\ 
SN 2003gk & NGC 7460    & Ib        & 44 & 6.0 & $2260 \pm 130$ \\ 
SN 2003hp & UGC 10942   & Ic/BL     & 93 & 5.9 & $<201$ \\ 
SN 2003id & NGC 895     & Ic pec    & 30 & 5.7 & $<278$ \\ 
SN 2003ig & UGC 2971    & Ic        & 79 & 5.8 & $<244 \phn (106 \pm 46)$ \\ 
SN 2003ih & UGC 2836    & Ib/c      & 68 & 5.7 & $<120$ \\ 
SN 2003jd & NGC 132     & Ic/BL     & 77 & 5.7 & $<172$ \\ 
SN 2003jg & NGC 2997    & Ib/c      & 13 & 5.7 & $<330$ \\ 
SN 2004ao & UGC 10862   & Ib        & 30 & 5.4 & $<256$ \\ 
SN 2004aw & NGC 3997    & Ic/BL     & 73 & 5.1 & $<211$ \\ 
SN 2004bm & NGC 3437    & Ic        & 24 & 5.0 & $<314$ \\ 
SN 2004bf & UGC 8739    & Ic        & 77 & 5.3 & $<460 \phn (154 \pm 102)$ \\ 
SN 2004bs & NGC 3323    & Ib        & 77 & 5.2 & $<246$ \\ 
SN 2004bu & UGC 10089   & Ic/BL     & 84 & 5.0 & $<244$ \\ 
SN 2004dn & UGC 2069    & Ic        & 51 & 4.9 & $<196$ \\ 
SN 2004fe & NGC 132     & Ic        & 72 & 4.6 & $<278$ \\ 
SN 2004ge & UGC 3555    & Ic        & 67 & 4.6 & $<180$ \\ 
SN 2004gt & NGC 4038    & Ic        & 23 & 4.5 & $<593$\tablenotemark{f} \\ 
SN 2004gv & NGC 856     & Ib        & 79 & 4.5 & $<280 \phn (106 \pm 58)$\\ 
SN 2005E  & NGC 1032    & Ib/c      & 36 & 4.4 & $<267$ \\ 
SN 2005N  & NGC 5420    & Ib/c      & 76 & 4.8 & $<285$ \\ 
SN 2005V  & NGC 2146    & Ib/c      & 17 & 4.4 & $<750$\tablenotemark{f} \\ 
SN 2005aj & UGC 2411    & Ic        & 38 & 4.4 & $<143$\\ 
SN 2005ct & NGC 207     & Ic        & 54 & 4.0 & $<327$\\ 
SN 2005da & UGC 11301   & Ic/BL     & 68 & 3.9 & $<309$\\ 
SN 2005dg & ESO 420-3   & Ic        & 56 & 3.9 & $<370 \phn (121\pm83)$\\ 
SN 2005ek & UGC 2526    & Ic        & 67 & 3.7 & $<171$\\ 
SN 2005eo & UGC 4132    & Ic        & 74 & 3.8 & $<200$\\ 
SN 2005kz & MCG+08-34-32& Ic/BL     & 115& 3.6 & $<242$\\ 
SN 2005lr & ESO 492-02  & Ic        & 36 & 3.5 & $<185$\\ 
SN 2006F  & NGC 935     & Ib        & 55 & 3.5 & $<268$\\ 
SN 2006ab & PGC 10652   & Ic        & 68 & 3.3 & $<159$\\ 
SN 2006dg & IC 1508     & Ic        & 58 & 3.2 & $<180$\\ 
SN 2006dj & UGC 12287   & Ib        & 73 & 3.2 & $<195$\\ 
SN 2006eg & CGCG462-023 & Ibc       & 53 & 2.9 & $<214$\\ 
SN 2006ep & NGC 214     & Ib        & 61 & 2.7 & $<284$\\ 
SN 2007D  & UGC 2653    & Ic/BL     & 93 & 2.5 & $<171$\\ 
SN 2007Y  & NGC 1187    & Ib        & 18 & 2.3 & $<177$\\ 
SN 2007iq & UGC 3416    & Ic        & 57 & 1.9 & $<126$\\ 
SN 2007ke & NGC 1129    & Ib        & 70 & 1.7 & $<146$\\ 
SN 2007ru & UGC 12381   & Ic/BL     & 64 & 1.5 & $<232$\\ 
SN 2007rz & NGC 1590    & Ic        & 52 & 1.6 & $<170$\\ 
SN 2008du & NGC 7422    & Ic        & 66 & 0.9 & $<412 (136 \pm 92)$\\
SN 2008dv & NGC 1343    & Ic        & 33 & 1.0 & $<450$\\ 
\end{tabular} 
\\
\tablenotetext{a}{The SN type, taken from \citet{Barbon+2010},
  with ``BL'' indicating a broadline SN}
\tablenotetext{b}{The age of the SN, since estimated shock
  breakout or detection the date of observation, on 2009 May 25}
\tablenotetext{c}{The distance to the SN, derived from the
NED database}
\tablenotetext{d}{The observed 8.5~GHz flux density and its
uncertainty, or the 3$\sigma$ upper on it} 
\tablenotetext{f}{The flux density from the SN has been corrected
for significant radio emission from the galaxy
at the location of the SN} 
\end{minipage}
\end{table*}

\begin{figure*}
\centering
\includegraphics[width=0.7\linewidth]{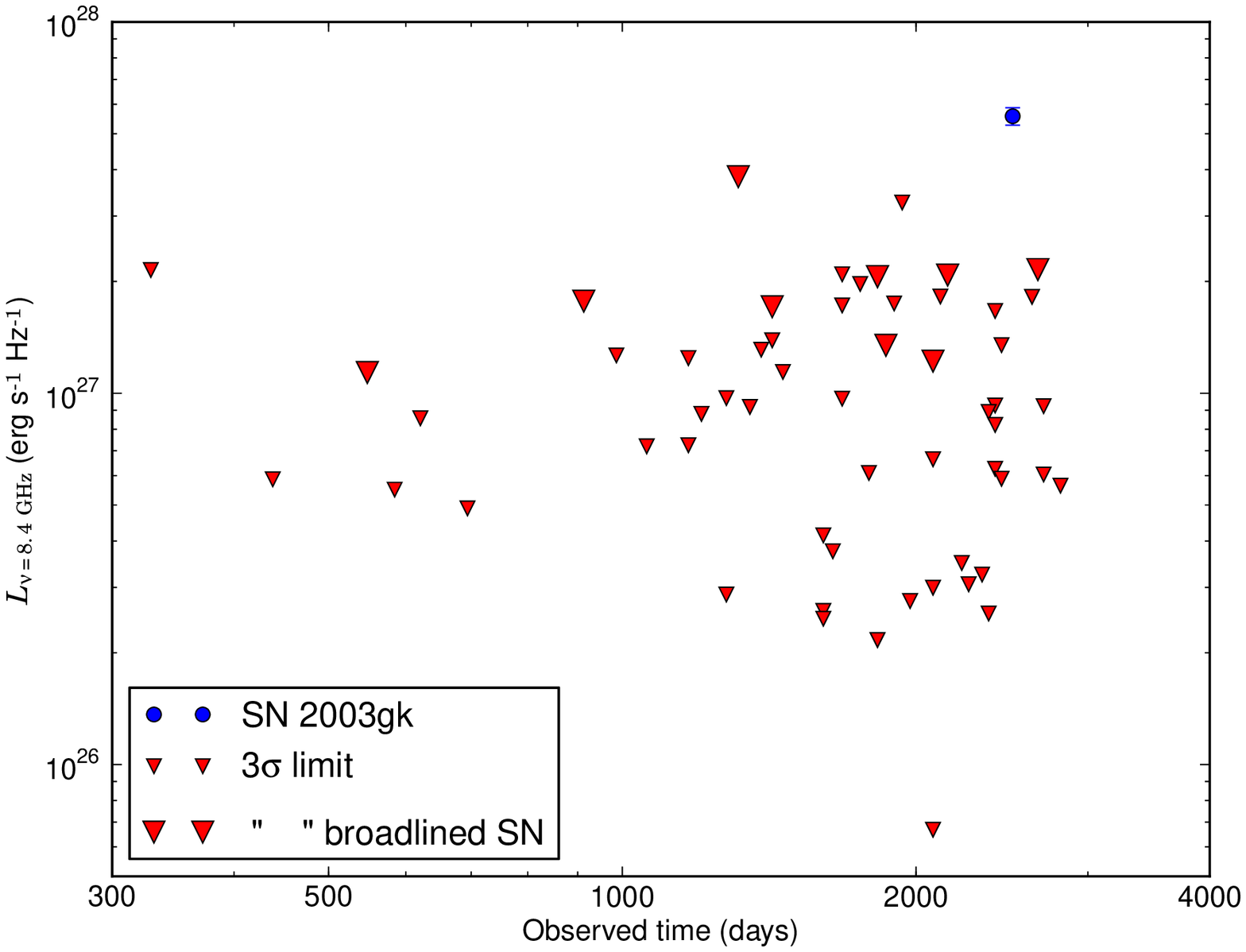}
\caption{The detection of, and upper limits on the late-time radio
  emission of Type b/c SNe.  We plot the 8.5~GHz spectral
  luminosity of the sole detection in our sample, SN~2003gk, in blue
  (note that the error bars are comparable in size to the plotted
  symbol), and the $3\sigma$ upper limits for the remaining 58 SNe as
  red triangles.  Broadlined SNe are marked with larger
  triangles.}
\label{flateradio}
\end{figure*}

From our measurements and limits to the flux density, we calculate the
corresponding values of or limits on the radio spectral luminosity at
8.5~GHz using the distances indicated in Table~\ref{tsne}.  We plot
these values in Figure~\ref{flateradio}.  The luminosity of SN~2003gk
was $(5.6 \pm 0.3) \times 10^{27}$~\ergsHz\ (where the uncertainty
does not include any uncertainty in the distance, which was taken to
be 44~Mpc).

We note that our survey is similar to that of \citet{Soderberg+2006b},
who also obtained upper limits on the 8.5-GHz flux density and thus
radio luminosity of Type I b/c SNe at late times.  We in fact
re-observed fourteen SNe from that earlier survey: SN 2001ej,
SN 2001is, SN 2002J, SN 2002bl, SN 2002cp, SN 2002hf, SN 2002ho, SN
2002hy, SN 2002hz, SN 2002ji, SN 2002jj, SN 2002jp, SN 2003dr, and SN
2003jd.  Our flux-density limits were broadly similar to those of
\citet{Soderberg+2006b}, but our observations occurred about 5.5 years
later than theirs, and thus set upper limits on a much later part of
the lightcurve.

\section{SN 2003gk}
\label{s2003gk}

\subsection{Additional VLA observations and radio lightcurve of SN 2003gk}

\objectname{SN~2003gk} was the only supernova detected in our radio
survey.
It was discovered by the Katzmann Automatic Imaging Telescope (KAIT)
on 2003 July 1.5 (UT) with an unfiltered magnitude of 17
\citep{GrahamL2003a, GrahamL2003b}.  Nothing was seen at its location
on a KAIT image from 2002 Dec.\ 3.2 to magnitude $\sim$19, and an
optical spectrum by \citet{Matheson+2003} showed it to be probably of
Type Ib, resembling SN~1984L several weeks after maximum light,
suggesting an explosion date around 2003 June 01 (MJD = 52792), which
date we adopt here.  \citet{Sollerman+2003},
deduced a relatively low expansion velocity of $\sim$8300~\kms\ from
the minimum of the He~I 587.6-nm absorption trough.  The SN occurred
in the Sc galaxy NGC 7460, which is at a distance of 45 Mpc
\citep[HyperLeda,][]{Paturel+2003}\footnote{The distance is derived
from the measured radial velocities, corrected for the local cluster's
infall velocity to Virgo; obtained from the HyperLeda database at {\tt
http://\-leda.univ-lyon1.fr}.}
In order to confirm our radio detection and to determine a light-curve
and a measure the radio spectral index, we obtained two additional
short VLA observations.  The first followup observation was on 2010
May 5 (observing code AB1351), where we observed at 8.46 and 22.5~GHz
with a total bandwidth of 512 MHz, and the array in the D
configuration.  The second was on 2012 May 30, using the VLA wideband
system, and we observed at 3, 8, 21~GHz, with a total bandwidth of
2~GHz.  In both cases we used 3C~48 as a flux density-density
calibrator (using the Perley-Butler 2010 coefficients), and we used
J2257+0243 as a phase and delay calibrator, and we used referenced
pointing at 18~GHz and higher.  The 2010 data set was reduced using
AIPS, while the 2012 data set was reduced using CASA
\citep{McMullin+2007}.

\begin{figure}
\centering
\includegraphics[width=\linewidth]{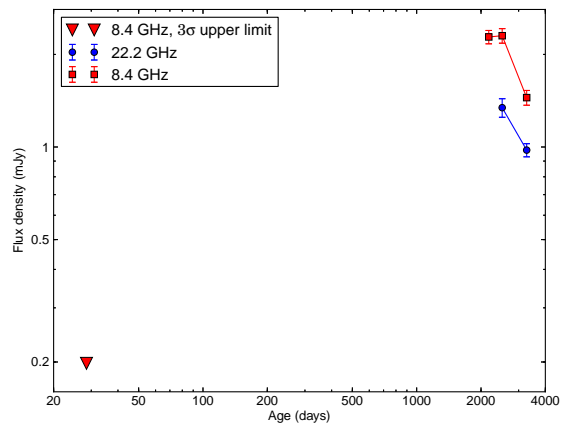}
\caption{The radio lightcurves of SN~2003gk at 8.4 and 22~GHz.
We plot the measured values with uncertainties as well as
the upper limit obtained from the early non-detection in 2003.
}
\label{flightcurve}
\end{figure}

We collect the flux density measurements of SN~2003gk in
Table~\ref{tlightcurve} and plot them in Figure~\ref{flightcurve}.
The mean value of the spectral index between 22.5 and 8.5~GHz
was $-0.5 \pm 0.1$.

\begin{table*}
\begin{minipage}[t]{\textwidth}
\caption{Flux Density Measurements of SN 2003gk}
\label{tlightcurve}
\begin{tabular}{l c c l}
Date & MJD & Frequency & Flux Density\tablenotemark{a} \\
     &  & (GHz) & $\mu$Jy \\
\hline
2003 07 14  & 52834  & \phn 8.46 & $< 1980$\tablenotemark{b} \\
2009 05 29  & 54981  & \phn 8.46 & $2280 \pm  110$ \\
2010 05 02  & 55318  & \phn 8.46 & $2300 \pm  130$ \\
2010 05 02  & 55318  &     22.46 & $1360 \pm  90$ \\
2012 05 30  & 56077  & \phn 8.46 & $1450 \pm  80$ \\
2012 05 30  & 56077  &     21.36 & $0960 \pm  50$ \\
\hline
\end{tabular}
\\
\tablenotemark{a}{The listed uncertainties include an assumed 5\% uncertainty
in the flux density calibration.} \\
\tablenotetext{b}{$3\sigma$ upper limit.} \\
\end{minipage}
\end{table*}

\begin{figure}
\centering
\includegraphics[width=\linewidth]{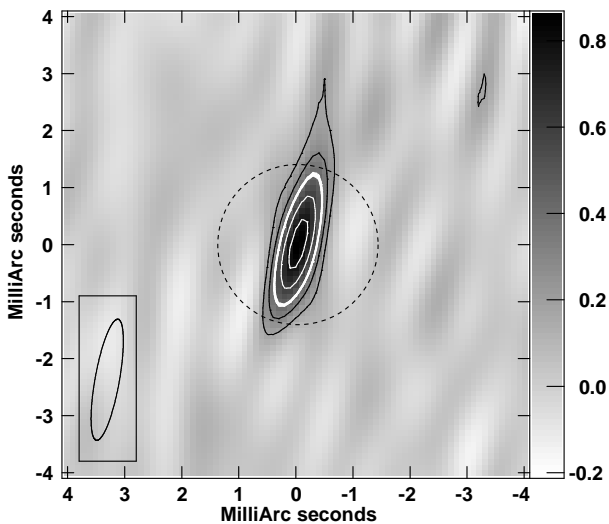}
\caption{A VLBI image of SN2003gk taken on 2011 Apr 21\@.  The
  contours are drawn at $-20$, 20, 30, {\bf 50}, 70 and 90\% of the peak
  brightness, which was 86~$\mu$\Jb.  The rms background brightness
  was 51~$\mu$\Jb.  The FWHM restoring beam of 2.16~mas $\times$
  0.43~mas is indicated at lower left.  North is up and east is to the
  left, and the coordinate origin is the brightness peak of SN~2003gk,
  which was at RA = \Ra{23}{01}{42}{98207}, dec =
  \dec{02}{16}{08}{6798} (J2000).  The dashed circle shows a circle of
  one light-year radius, showing the expected size of a source which
  has expanded relativistically for one year.}
\label{fvlbiimg}
\end{figure}

\subsection{VLBI observations}
\label{svlbiobs}

In order to determine the size of the radio emitting region in
SN~2003gk, and thus to determine its average expansion speed, we
obtained 8.4-GHz VLBI imaging observations of it on 2011 April 21
(observing code BB296), with a total time of 5 hours.  The midpoint of
the observations was at MJD 55673.  We used the High-Sensitivity
Array, which consisted of the NRAO VLBA ($8 \times 25$-m diameter; the
Pie Town and North Liberty antennas did not take part in this run),
the NRAO Robert C. Byrd $\sim$105~m telescope at Green Bank, the
Effelsberg (100~m diameter)\footnote{The telescope at Effelsberg is
operated by the Max-Planck-Institut f\"{u}r Radioastronomie in Bonn,
Germany.}  telescope and the Arecibo\footnote{The Arecibo Observatory
is part of the National Astronomy and Ionosphere Center, which is
operated by Cornell University under a cooperative agreement with the
National Science Foundation.} (305-m diameter) telescopes.

We recorded a bandwidth of 64~MHz in both senses of circular
polarization with two-bit sampling, for a total bit rate
512~Mbit~s$^{-1}$.
The VLBI data were correlated with NRAO's VLBA processor, and the
analysis carried out with NRAO's Astronomical Image Processing System
(AIPS)\@.  The initial flux density calibration was done through
measurements of the system temperature at each telescope, and then
improved through self-calibration of the reference source.  A
correction was made for the dispersive delay due to the ionosphere
using the AIPS task TECOR, although the effect at our frequency is not
large.

We phase-referenced our VLBI observations to \objectname{QSO
  J2257+0243}, which is an ICRF source for which we use the position
RA = \Ra{22}{57}{17}{563103}, dec.\ = \dec{02}{43}{17}{51172} (J2000)
\citep{Fey+2004}.
We used a cycle time of $\sim$3.9~min, with $\sim$2.4~min spent on
SN~2003gk.  We discarded any SN~2003gk data taken at elevations below
10\arcdeg.  In addition, we also spent two periods of $\sim$10~min
observing an astrometric check source \objectname{JVAS J2258+0203},
phase-referenced to J2257+0243 in the same manner as SN~2003gk.

We found that on both our check source, JVAS J2258+0203, and for
SN~2003gk, the visibility phases for baselines involving Arecibo (AR)
showed large residuals, suggesting that phase referencing at AR was
not successful, and we therefore did not use the AR data for any
astrometric results.

For marginally resolved sources, such as SN~2003gk, the best values
for the source size come from fitting models directly to the
visibility data, rather than from imaging.  We chose as a model the
projection of an optically-thin spherical shell of uniform volume
emissivity, with an outer radius of $1.25\times$ the inner
radius\footnote{Our results do not depend significantly on the assumed
  ratio between inner and outer radii, as the effect of reasonable
  variations in this ratio is considerably less than our stated
  uncertainties.  For a discussion of uncertainties on the shell-size
  obtained through \uv~plane modelfitting compared with those obtained
  in the image plane for the case of SN~1993J, showing that superior
  results are obtained using the former, see
  \citet{SN93J_Manchester}.}.
Such a model has been found to be appropriate for other radio
SNe \citep[see e.g.][]{SN93J-3, SN79C-Shell}.  The Fourier
transform of this shell model is then fitted to the visibility
measurements by least squares.

We obtained a value of 0.37~mas for the outer angular radius of
SN~2003gk.
We also fitted the same shell model, but added antenna amplitude gains
(non time-dependent scale factors) as free parameters, which
changed the fitted outer radius by +0.06~mas.  Since the
signal-to-noise ratio is too low to allow reliably fitting antenna
gains (a form of amplitude self-calibration) we keep the original
value of 0.37~mas as our best fit value, but take as a conservative
uncertainty the 0.06~mas difference between the value obtained with
the antenna gains added as free parameters and the original one.  This
value is approximately twice as large as the purely statistical
uncertainty.  We therefore take the final fitted value of the outer
angular radius of SN~2003gk as $0.37 \pm 0.06$~mas.

For a partially resolved source such as SN~2003gk, the exact model
geometry is not critical, and our shell model will give a reasonable
estimate of the size of any circularly symmetric source, with a
scaling factor of order unity dependent on the exact morphology
\citep[see discussion in][]{SN93J-2}.  In particular, using a circular
Gaussian model instead of the spherical shell model would result in a
fitted FWHM size of 0.5~mas, with the same relative uncertainty of
16\%.  We also attempted a fit of an elliptical Gaussian to model a
possibly elongated source.  To reduce the number of free parameters we
fixed the axis ratio to 0.2\@.  We obtained a FWHM major axis size of
$0.61 \pm 0.10$~mas.

Our fitted angular outer radius for SN~2003gk was $0.37 \pm 0.06$~mas
(for a spherical shell model).  At a distance of 44~Mpc, this
corresponds to $(2.4 \pm 0.4) \times 10^{17}$~cm.  The age of the
SN at the time of the VLBI observations was 2881~days, so the
average expansion velocity was $(1.0 \pm 0.2) \times 10^4$~\kms.  The
measured radius is not compatible with any relativistic, or
near-relativistic expansion: at an apparent speed of $c$ it would have
reached our measured size at $t = 96$~d, so any reasonable
non-relativistic expansion speed in the $\sim$7.5 yr since then would
have increased the size well beyond our measured value.  The measured
size is, however, entirely compatible with the expansion velocities of
ordinary, non-relativistic SNe \citep[e.g.][]{VLBA10th,
  Bartel2009}.  Our VLBI measurements, therefore, exclude
any relativistically expanding jet component in SN~2003gk.
Unfortunately, with only single epoch of VLBI observations,
we cannot constrain the proper motion.


\section{Calculation of model lightcurves}
\label{smodels}

The association of GRBs with massive stars implies that the afterglow
shock propagates into the pre-explosion stellar wind, and suggests a
stratified external medium with a density profile $\rho_{\rm ext} = A
r^{-k}$. If the ratio of wind velocity, $v_w$, to mass-loss rate,
$\dot{M}_w$, remains constant, then $k = 2$ and $A =\dot{M}_w/(4\pi
v_w) = 5\times 10^{11}A_*\;{\rm g\;cm^{-1}}$. Since $\dot{M}_w/v_w$
might vary before the SN explosion, and is rather uncertain,
other values of $k$ have also been considered both in modelling of GRB
afterglows
\citep[e.g.][]{Yost03,Starling08,Leventis+2012,Leventis+2013} and
recently also in hydrodynamic simulations \citep{DeColle+2012b}.

Nonetheless, most afterglow lightcurves calculated so far, and in
particular those from hydrodynamic simulations, have been done for a
uniform external medium ($k=0$). Therefore, we present here results
for a wind-like external medium of constant $\dot{M}_w/v_w$
(i.e.\ $k=2$), as a representative value for what might more
realistically be expected for the wind of a massive star progenitor.

The typical value of the external density normalization parameter,
$A_*$, is usually taken to be 1.0 in modelling, although the values
that are inferred from afterglow broadband modelling range from $A_*
\simeq 1$ down to less than 0.01.
\citep{KP03,Waxman2004a,Waxman2004b,ChevalierLF2004,Rol07,
  Racusin+2008,Pandey09,Cenko10,Cenko11}.

The true jet kinetic energy is usually inferred to be $E_{\rm jet}
\sim 10^{50}-10^{51.5}$~erg, for bright well-monitored afterglows and
up to $\sim 10^{52}$~erg for the most energetic afterglows
\citep[e.g.][]{PK01a,PK01b,Yost03,Cenko10,Cenko11}.  However,
low-luminosity GRBs that have a larger rate per unit volume extend
this distribution down to $\lesssim 10^{48}$~erg
\citep[e.g.][]{Hjorth2013}.

The shock-microphysics processes responsible for field amplification
and particle acceleration are typically parametrized by the
assumptions that the magnetic field everywhere in the shocked region
holds a fraction $\epsilon_B = 0.1$ of the local internal energy
density in the flow, and that the non-thermal electrons just behind
the shock hold a fraction $\epsilon_e = 0.1$ of the internal energy
and have a power-law energy distribution with $N(E) \propto E^{-p}$.
When it is possible to infer the values of these microphysics
parameters for particular bursts, they are typically in the ranges
$10^{-5}\lesssim\epsilon_B\lesssim 10^{-1}$,
$10^{-2}\lesssim\epsilon_e\lesssim 10^{-0.5}$ and $2\lesssim p\lesssim
3$ \citep[e.g.][]{Santana13}.  We refer to the values $E_{\rm jet} = 2
\times 10^{51}$~erg, $A_* = 1, \epsilon_B = \epsilon_e = 0.1$, which
are often used in modelling, as the ``canonical'' values, although, as
just mentioned, they are likely not representative of the majority of
bursts.

We use 2D hydrodynamic simulations for $k=2$ from
\citet{DeColle+2012b}, based on the special relativistic hydrodynamics
code \emph{Mezcal}, and a complimentary code for calculating the
radiation by post-processing the results of the numerical simulations
\citep{DeColle+2012a}.  The GRB was initialized on a conical wedge of
half-opening angle $\theta_0=0.2$~rad, taken out of the spherical
self-similar \citet{BlandfordM1976} solution. The simulation starts
when the Lorentz factor of the material just behind the shock was
$\Gamma=20$. The calculation of the synchrotron radiation is
supplemented by adding the contribution from a \citet{BlandfordM1976}
conical wedge at earlier times, corresponding to $20\leq\Gamma\leq
500$ (which causes an artificially sharp transition in the lightcurve
between the two at a rather early time). Our value of $\theta_0=0.2$
corresponds to a beaming factor of $f_b = 1-\cos\theta_0\approx
0.02$. The simulation was for $\Ekiso = 10^{53}$~erg, corresponding to
$E_{\rm jet} = f_b \Ekiso \approx 2\times 10^{51}$~erg, and for $A_* =
1.65$, but was scaled to arbitrary values of $E_{\rm jet}$ and $A_*$
using appropriate scaling relations from \citet{Granot2012}. We have
fixed the power-law index of the accelerated electrons to $p=2.5$ as a
representative value.

\begin{figure*}
\centering
\includegraphics[width=0.7\linewidth]{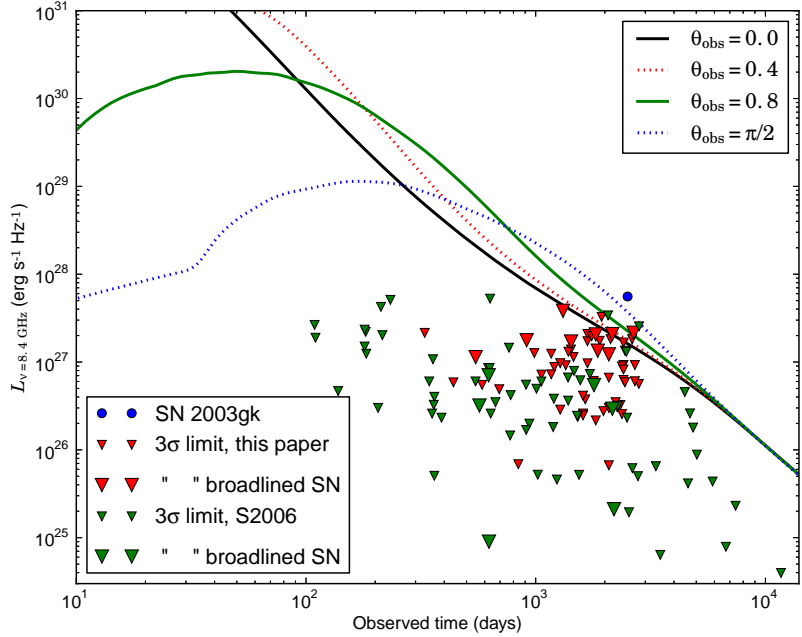}
\caption{The modelled 8.4-GHz lightcurves of relativistic jets for
  angles to the line of sight between 0 and $\pi/2$ radians, as
  indicated at top right.  The lightcurves were calculated assuming a
  wind-stratified medium ($k=2$) with $A_* = 1$ and $\epsilon_B =
  \epsilon_e = 0.1$ (see text \S~\ref{smodels}).  The red triangles
  our upper limits while the blue point is the measured values for
  SN~2003gk, repeated from Figure~\ref{flateradio} above.  We add here
  the corresponding limits from \citet[][marked
    ``S2006'']{Soderberg+2006b} as green triangles.  The larger
  triangles again represent broadlined SNe.  Note that for 14 SNe,
  there are two separate limits, one from our observations and an
  earlier one from \citet{Soderberg+2006b}.  Although some of our
  limits (red triangles) are above the modelled lightcurves, in each
  such case there is an earlier limit (green triangle) which is well
  below the modelled curves.}
\label{fnominal}
\end{figure*}

\begin{figure*}
\centering
\includegraphics[width=0.7\linewidth]{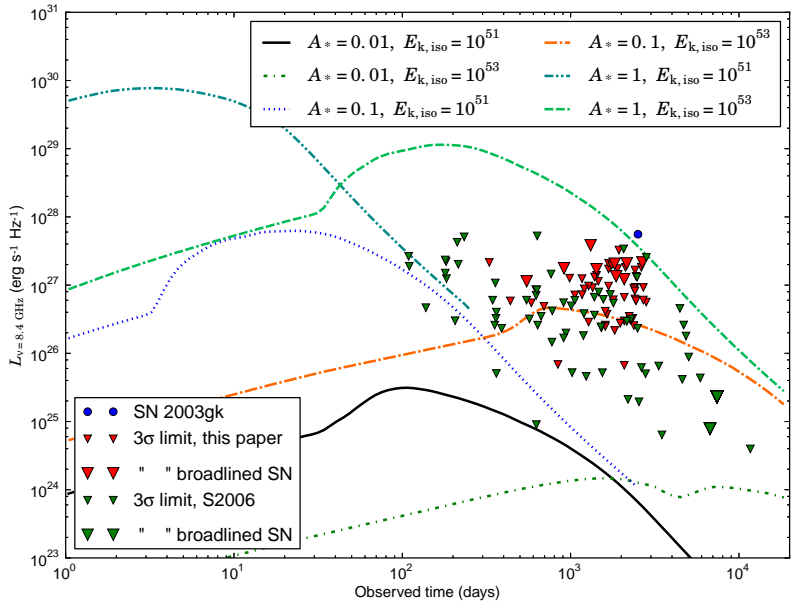}
\caption{The modelled lightcurves for various possible explosion
energies and circumstellar densities, all for and angle to the line of
sight $\thobs = \pi/2$.  The curves are for the indicated values of
\Ekiso, the isotropic explosion energy in erg and for a circumstellar
density parameter $A_*$.  For comparison, we again plot the observed
value for SN~2003gk and limits for the other SNe from our sample (see
Figure~\ref{fnominal}).
}
\label{fenergyetc}
\end{figure*}

\begin{figure*}
\centering
\includegraphics[width=0.7\linewidth]{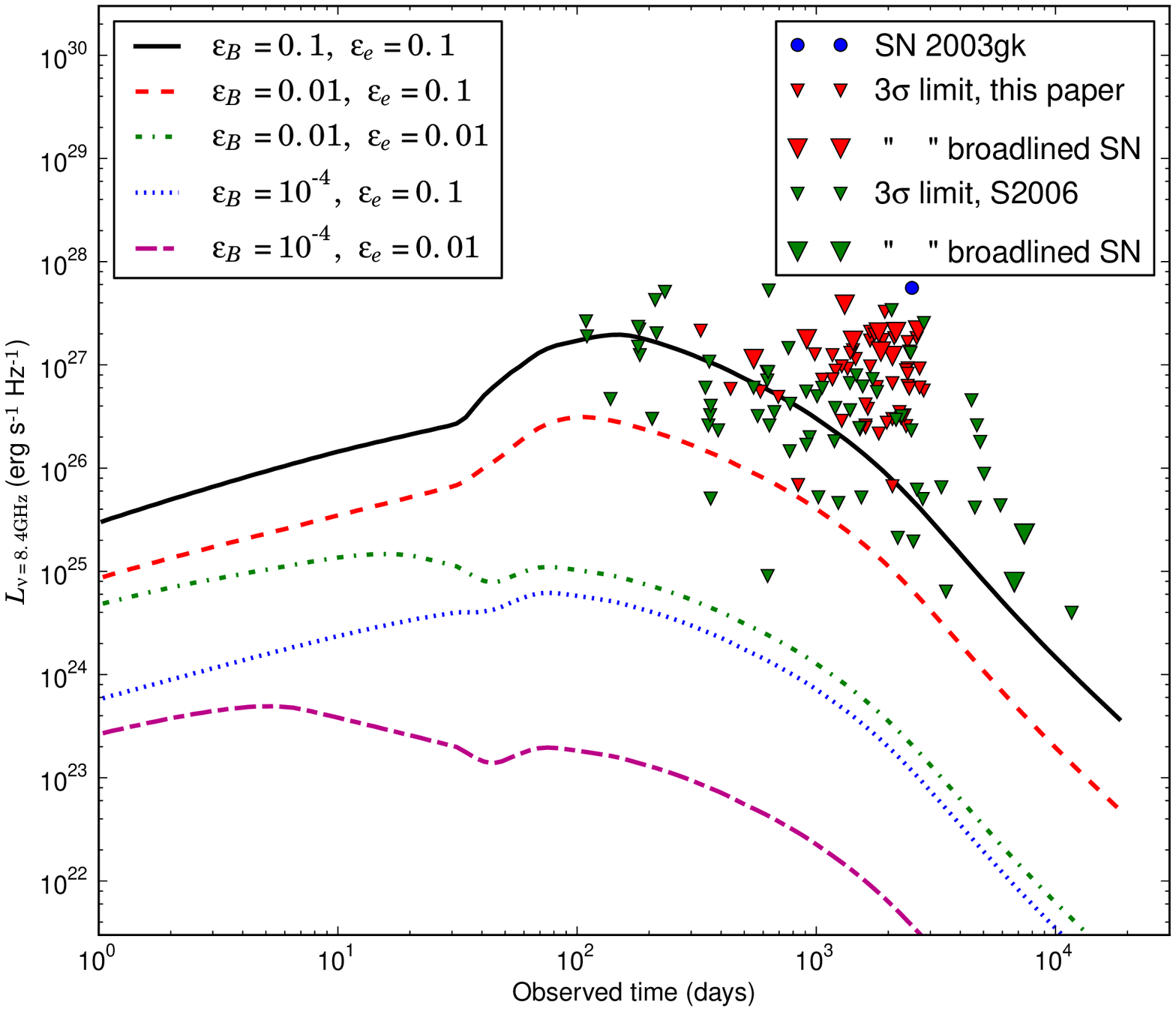}
\caption{The modelled lightcurves for various possible efficiencies of
  magnetic field generation ($\epsilon_B$) and particle acceleration
  ($\epsilon_e$) at the shock front.  All the lightcurves are for a
  wind-stratified medium ($k = 2$), $A_* = 0.1$, $\Ekiso =
  10^{52}$~erg and for $\thobs = \pi/2$.  The lightcurves are shown
  for the indicated values of ($\epsilon_B$ and $\epsilon_e$).  For
  comparison, we again plot the observed value for SN~2003gk and
  limits for the other SNe from our sample (see
  Figure~\ref{fnominal}).}
\label{fmicrophysics}
\end{figure*}

Fig.~\ref{fnominal} shows lightcurves for different viewing angles
($\thobs = 0,\,0.4,\,0.8,\,\pi/2$) for our optimistic model using the
canonical parameters which produce relatively bright radio afterglow
emission: $\epsilon_B = \epsilon_e = 0.1$, $E_{\rm k,iso} =
10^{53}$~erg, and $A_*=1$.  In Fig.~\ref{fenergyetc} we fix
$\epsilon_B = \epsilon_e = 0.1$ (as well as $\thobs=\pi/2$) and show
the effect of varying the jet energy ($\Ekiso =10^{51},\,10^{53}$~erg)
and the external density normalization ($A_*=0.01,\,0.1,\,1$), to
cover a range that might be considered more typical for GRB
jets. Hydrodynamics features, such as the observed non-relativistic
transition time (which typically corresponds to the peak of the
lightcurve for large off-axis viewing angles) scales as
$(\Ekiso/A)^{1/(3-k)}$, and therefore they vary much more strongly for
the wind-like external medium (as $\Ekiso/A_*$ for $k=2$) than for a
uniform medium where they vary only as $(\Ekiso/n_{\rm
  external})^{1/3}$.  Not only the peak time varies substantially, but
also the peak flux, which depends very strongly on
$A_*$\@. Fig.~\ref{fmicrophysics} fixes $\Ekiso =10^{52}$~erg and
$A_*=0.1$ (as well as $\thobs=\pi/2$), and shows the dependence of the
lightcurves on $\epsilon_B$ and $\epsilon_e$ when varying the latter
two well within the typical range inferred from GRB afterglow
observations ($10^{-4}\leq\epsilon_B\leq 0.1$ and
$10^{-2}\leq\epsilon_B\leq 0.1$). This variation also has a large
effect on the flux-density normalization.

The peak time of the modelled lightcurves depends on $(\Ekiso/A_*)$
and with larger values of this ratio producing later peaks.  The
lightcurve with the canonical values of $\Ekiso = 10^{53}$~erg and
$A_* = 1$ peaks at $t = 109$~d with 8.4-GHz spectral luminosity,
\LXband = $1.1 \times 10^{29}$~\ergsHz, while the faintest of the
lightcurves, also with $\Ekiso = 10^{53}$~\ergsHz\ but with $A_* =
0.01$ peaks at $t = 1466$~d and \LXband = $1.5 \times
10^{24}$~\ergsHz.  The lightcurve with $\Ekiso = 10^{51}$~erg and $A_*
= 1$ peaks as early as 3~d with \LXband = $8 \times 10^{29}$~\ergsHz.
In addition, if either $\epsilon_B$ or $\epsilon_e$ are below the
nominal values of 0.1, a fainter lightcurve results, and the delayed
peak, which is characteristic of jets at large angles to the line of
sight, becomes less prominent.  This occurs since the peak frequency,
$\nu_{\rm m} \propto\epsilon_B^{1/2}\epsilon_e^2$, and therefore is
lower and passes the frequency of observation earlier, before the time
when the beaming cone of the jet's radiation reaches our line of sight.

\section{Comparison of observed limits to model lightcurves}
\label{scompare}

We now compare the model lightcurves for relativistic jets at various
angles to the line of sight to the measurements of the late time radio
emission of Type I b/c SNe.  We combine our own sample (section
\ref{ssurveyresults} above) with that of \citet{Soderberg+2006b}.  Our
combined sample consists of 126 upper limits on 112 different SNe (14
SNe have limits obtained at two different times).  We exclude
SN~2003gk from this discussion, because as we have shown, its radio
emission is not due to a relativistic jet.

The model lightcurves are strongly dependent on the explosion energy,
\Ekiso\ and the circumstellar density normalization, $A_*$. We first
adopt ``canonical'', or optimistic, values of $\Ekiso = 10^{53}$~erg
and $A_* = 1$, and show the resulting lightcurves, along with the
observed upper limits, in Figure~\ref{fnominal}.  If we assume that
the jets are randomly oriented, and that the statistics are Gaussian,
we can calculate the probability ($P$) that any such jet would be fall
below the limits we measured for our sample of SNe.  The SN most
compatible with the canonical lightcurves by this criterion is the
non-broadlined SN~1996D, with $P \simeq 0.014$, while for broadlined
SNe it is SN~2003hp with $P \simeq 10^{-4}$.
We can therefore conclude that the probability of {\em any} of 112 our
SNe being as bright as our canonical lightcurves is $<2$\%, and that
probability that any of the 13 broadlined SNe being as bright is $<
10^{-3}$.  Note that a few of the limits from this paper {\em are}
above the model lightcurves, but only for SNe for which an earlier
limit for the same SN from \citet{Soderberg+2006b} was well below the
model lightcurves (as noted, we exclude SN~2003gk here, which is well
above the predicted lightcurves, but as we showed above does not have
any relativistic jet).

The brightness of the modelled lightcurves, however, depends strongly
on the explosion energy (\Ekiso) and the density of the circumstellar
medium ($A_*$).  The canonical values above were adopted for {\em
observed} GRB afterglows, and almost certainly represent present
particularly bright GRB jets rather than the typical ones.  In
Figure~\ref{fenergyetc}, we therefore show the model lightcurves for a
variety of plausible values for \Ekiso\ and $A_*$.

As can be seen, regardless of the value of $A_*$, all the lightcurves
with $\Ekiso = 10^{51}$~erg are compatible with most of our observed
limits, and even the lightcurves with the canonical value of $\Ekiso =
10^{53}$~erg are compatible with most of our limits provided that the
CSM density is characterized by $A_* \lesssim 0.1$.  Even for the canonical
values of $\Ekiso = 10^{53}$~erg and $A_* = 1$, the predicted
lightcurves fall below many of our measured limits if either
$\epsilon_B$ or $\epsilon_e$ is below the canonical value of 0.1, but
still within the range inferred to actually occur in GRBs.

\section{Discussion}
\label{sdiscuss}

\subsection{SN 2003gk}
\label{s2003gkdiscuss}

SN~2003gk was not detected in the radio early on, with an upper limit
to the 8.4-GHz spectral luminosity of $\LXband = 4 \times
10^{25}$~\ergsHz\ at $t = 29$~d.  By $t=2881$~d, the radio
luminosity had risen to $\sim4 \times 10^{26}$~\ergsHz, and appeared
to be decaying rapidly, with $\LXband \propto t^{1.7 \pm 0.3}$ between
$t = 2510$ and $3270$~d.  The spectral index between 8.4 and 22~GHz
was $-0.6 \pm 0.2$.

Our VLBI measurements (section \ref{svlbiobs}) showed that SN~2003gk
was expanding non-relativistically, with an average speed of $(1.0 \pm
0.2) \times 10^4$~\kms.  If we assume a power-law expansion, with
radius $\propto t^m$ and take a typical value of 0.8 \citep[see,
  e.g.][]{Weiler+2002} for the deceleration parameter, $m$,
then we can calculate that the initial speed ($t =
30$~d) was $20\:000$~\kms, which is within the range usually seen for
core collapse SNe in general.  Type I b/c SNe tend to have somewhat
higher speeds than other SNe, with e.g.\ \citet{ChevalierF2006}
listing a median speed of $\sim 42\:000$~\kms.  If we take the
expansion speed of SN~2003gk to have been $42\:000$~\kms\ at $t =
30$~d, we can calculate that SN~2003gk must have been fairly strongly
decelerated, with $m \simeq 0.6$.

We mentioned SN~2001em earlier, which showed a similar evolution in
flux-density, but for which the VLBI observations also implied only
non-relativistic expansion.  We propose for SN~2003gk an explanation
similar to that proposed for SN~2001em
\citep[see][]{ChugaiC2006,Chevalier2007}, namely that the radio
emission is produced by the interaction of a normal Type I~b/c ejecta
with a massive and dense circumstellar shell at some distance from the
progenitor
The shell was the result of episodic mass-loss from the progenitor,
perhaps from a luminous blue variable like eruptive event.  A possible
further diagnostic would be if SN~2003gk were to show strong H$\alpha$
emission, with a relatively narrow line width, which would be expected
to accompany the strong circumstellar interaction.

\subsection{What fraction of Type I b/c SNe host a GRB?}
\label{sfraction}

We carried out a survey for late-onset radio emission in Type I b/c
SNe that might be indicative of an off-axis relativistic jet.  Only
one of our 59 SNe, SN~2003gk, showed any such radio emission, but our
VLBI observations of it rule out relativistic expansion.  It is clear
therefore, that regardless of orientation, only a small fraction of
Type I b/c have a relativistic jet producing bright late-time radio
emission, or in other words host a GRB event.  This conclusion was
already reached earlier, in particular by
\citet{SoderbergFW2004,Soderberg+2006b}.
We have combined our present sample with that of
\citet{Soderberg+2006b}, for a combined set of 112 Type I b/c SNe
which
have been examined for radio emission such as might arise from an
off-axis relativistic jet.

We compared limits on radio emission obtained for this combined sample
to model lightcurves for relativistic jets at various angles to the
line of sight (section \ref{scompare}), using numerically modelled
lightcurves based on hydro-dynamic simulations, rather than the
semi-analytic ones of \citet{Soderberg+2006b}.

On the basis of our results, the hypothesis that all Type I b/c SNe
have radio lightcurves as bright as the canonical models with $\Ekiso
= 10^{53}$~erg, $A_* = 1$, and $\epsilon_B = \epsilon_e = 0.1$ can be
rejected with a high level of confidence.  Our sample included 13
broadlined Type I b/c SNe, and we can also reject the hypothesis that
all broadlined SNe have such bright radio lightcurves.  We performed
Monte-Carlo simulations with $10\:000$ trials, a randomly chosen
fraction $f_{\rm bright}$ of our sample of 112 SNe having lightcurves
as bright as our canonical models for each trial, with the remainder
being unobservably faint.  We then compared the simulated brightness
values to our observed values or limits, and calculated the
probability of that particular trial given the observed values and the
uncertainties (assuming a Gaussian distribution for the measurement
errors with the values of $\sigma$ given or implied by the listed
$3\sigma$ limit in Table~\ref{tsne}).  We performed such simulations
for various values of $f_{\rm bright}$, with the result that we can
say (at the 99\% confidence level) that fewer than 5\% (i.e.\ $f_{\rm
bright} < 0.05$) of all Type I b/c SNe, and fewer than 33\% of
broadlined SNe have radio lightcurves as bright as those produced by
our canonical models.
Our conclusions are consistent with those of \citet{Soderberg+2006b},
who concluded that at most 10\% of all Type I b/c SNe are associated
with ``typical'' GRB jets regardless of orientation, where their
``typical'' GRB jets have radio luminosities similar to those of the
canonical models.  \citet{Soderberg+2006b} further concluded that even
of the broadlined SNe, at most a fraction can be associated with a GRB
jet.

Our results are also in agreement with the conclusions
of \citet{Ghirlanda+2013a} who compared a simulated population of
GRBs\footnote{\citet{Ghirlanda+2013a}'s synthesized a population of
  GRBs under the assumption that, in the rest frame, all GRBs emit a
  total gamma-ray energy of $1.5 \times 10^{48}$~erg have an $\nu
  F_\nu$ spectrum that peaks at 1.5~keV.}
to samples of observed GRBs from {\em Swift, Fermi}~GBM and {\em CGRO}
BATSE (1177 GRBs in total).  They found that to match the observed
rates of bright GRB detections, the rate of GRB events (at any
orientation) was $\sim$0.3\% the rate of local Type I b/c SNe, and
$\sim$4.3\% that of local BL SNe.  Although their constraint on the
fraction of Type I b/c SNe accompanied by a bright GRB are tighter
than ours, the two estimates are complimentary, since the two
estimates have differing model dependencies.  In particular, the radio
observations are sensitive to jets with wide range of $\Gamma$, whereas
the observational constraints used by Ghirlanda et al.\ were
restricted to highly relativistic jets with $\Gamma \gtrsim 100$.

However, the conclusion that the absence of late-time radio emission
rules out off-axis GRB bursts in most type I b/c SNe are based on the
assumption of a fairly bright jet, comparable to the {\em detected}
GRB afterglows.  Many of the fainter GRB events will go un-detected
\citep{WandermanP2010, Shahmoradi2013}, and radio afterglows are seen
only in a fraction of the detected GRBs.  In particular, that
conclusion was based on the assumption of a relatively energetic burst
with $\Ekiso = 10^{53}$~erg, with a relatively dense CSM,
characterized by $A_* = 1$, and with a shock that efficiently
generates magnetic field and relativistic particles.  These
assumptions apply to the bright end of the detected radio
afterglows. and almost certainly overestimate the radio emission
produced by the average burst.  To estimate the frequency of
relativistic jets in Type I b/c SNe, we want to use values
representative of the population rather than of the bright detected
bursts.

Our model calculations (Section \ref{smodels}) in fact used a wide
range of values for the jet characteristics, and should be more
representative of the full range of radio brightnesses that might be
expected from relativistic jets.  They show that, for jets with lower
explosion energies, less dense CSM, or lower values of
$\epsilon_B$ or $\epsilon_e$, than the canonical values, the predicted
radio lightcurves can be several orders of magnitude lower in flux
density.  We find that the peak 8.4-GHz spectral luminosity, \LXband,
of the off-axis relativistic jet can range between $8 \times
10^{29}$~\ergsHz\ (for the canonical values of \Ekiso\ and $A_*$) and
$1.4 \times 10^{24}$~\ergsHz\ (for \Ekiso = $10^{51}$~erg and $A_* =
0.01$, corresponding respectively to 260 m\Jb\ to 0.5~\muJb\ at a
distance of 50~Mpc.  Note that even for energetic bursts with \Ekiso =
$10^{51}$~erg, the radio brightness can be several orders of magnitude
lower than the canonical one.

Although analytical studies and simulations for a constant-density CSM
predicted that radio emission from a jet in the plane of the sky would
peak at $t \simeq 500$~d \citep{GranotL2003, vEertenZM2010,
vEertenM2012}, we find that for a wind-like CSM, with $\rho_{\rm ext}
\propto r^{-2}$, the peak time depends strongly on the burst energy
and CSM density, with the dependence being as $\Ekiso / A_*$, compared
to $(\Ekiso / \rho_{\rm ext})^{1/3}$ for a uniform density CSM\@.  For
our range of values, the 8.5-GHz peak time could be as short as 3~d
(for $\Ekiso = 10^{53}$~erg and $A_* = 1$, or as long as $\sim$4~yr
(for $\Ekiso = 10^{53}$~erg and $A_* =0.01$) for $\theta_{\rm obs} =
\pi/2$ (for other viewing angles this time changes, as can be seen in
Fig.~\ref{fnominal}).
Thus the model lightcurves for off-axis relativistic jets overlap with
those seen for normal SNe without any relativistic ejecta, which have
peak values of \LXband\ up to $10^{28}$~\ergsHz\ and times to peak as
long as 100~d \citep[e.g.][]{Soderberg+2005, Soderberg+2006e}.
We note, however, that the radio emission from non-relativistic SNe
shows a marked change of spectral index near the peak time as the
SN transitions from the optically-thick to optically-thin
regimes, while emission from an off-axis relativistic jet often shows
no such change.  A peak in the radio flux density not marked by a
change in the spectral index from an optically thick value of ($S_\nu
\propto \nu^{5/2}$ or $\nu^{2}$) before the break, to an optically thin
value after, may therefore be diagnostic of off-axis jet emission.

We can conclude therefore that at present, late-time radio
observations of Type I b/c SNe, even the ones with broad lines such as
have been associated with GRBs, place only weak constraints on the
presence of any possible off-axis relativistic (GRB-like) jets
associated with these objects.  In particular, burst energies lower
than $\Ekiso = 10^{53}$~erg, low-density CSM and lower efficiency of
field amplification or particle acceleration at the shock would all
produce jets that have only faint radio emission below the present
observational limits.

\section{Conclusions}
\label{ssummary}

Here is a summary of our main conclusions:

\begin{trivlist}

\item{1.}  Late-time radio emission was detected from SN~2003gk, with
  an 8.4-GHz flux density of $2.30 \pm 0.13$~mJy at age $\sim$8~yr.
  No radio emission was present at the age of $\sim$1 month.  The
  late-time radio emission is decaying with time, and has a spectral
  index of $\-0.5 \pm 0.1$.

\item{2.}  VLBI observations of SN~2003gk showed that its average
  expansion speed at age $\sim$8~yr was $10\:000 \pm 2000$~\kms, which
  is incompatible with relativistic expansion over more than $t \simeq
  100$~d.  The speed is, however, comparable to what is seen in
  normal, non-relativistic SNe.  The radio emission is likely due to
  the SN shock impacting upon a dense shell in the
  circumstellar material of the progenitor.  A diagnostic would be if
  SN 2003gk were to show strong H$\alpha$ emission with a relatively
  narrow line-width.

\item{3.} We surveyed 58 other Type I b/c SNe for late-time radio
  emission, such as might be produced by an off-axis GRB jet. We
  combined our results with those of similar, earlier survey by
  \citet{Soderberg+2006b}, for a total sample of 112 Type I b/c SNe.
  None of the 112 Type I b/c SNe surveyed, including 13 broadlined
  SNe, showed any radio emission attributable to a off-axis GRB jet.

\item{4.} We calculated new model lightcurves for relativistic jets in
  a wind-stratified (density $\propto r^{-2}$) circumstellar medium
  for various angles to the line of sight and for various values of
  the energy of the burst, the circumstellar density, and the
  efficiency of magnetic-field amplification and particle acceleration
  at the shock front.

\item{5.} For canonical parameters, such as are typical of detected GRBs,
  the predicted radio lightcurves for jets in the plane of the sky are bright, with
  peak 8.4-GHz spectral luminosities of $8\times10^{29}$~\ergsHz.
  However, varying the four parameters above within reasonable ranges
  can reduce the brightness by several orders of magnitude.

\item{6.}  Based on our simulations, the radio lightcurves of off-axis
  relativistic jets overlap with those or ordinary, non-relativistic
  SNe both in peak luminosity and time to reach the maximum, even for
  jets in the plane of the sky.  The present radio observations
  therefore do not place any strong constraints on the presence of
  off-axis jets in Type I b/c SNe.

\end{trivlist}

\section*{Acknowledgements}
\noindent{Research at Hartebeesthoek Radio Astronomy Observatory was
  partly supported by National Research Foundation (NRF) of South
  Africa.  Research at York University was partly supported by
  NSERC\@.  FDC aknowledges support from the DGAPA-PAPIIT-UNAM grant
  IA101413-2. The National Radio Astronomy Observatory (NRAO) is a
  facility of the National Science Foundation operated under
  cooperative agreement by Associated Universities, Inc.  The Arecibo
  Observatory is operated by SRI International under a cooperative
  agreement with the National Science Foundation (AST-1100968), and in
  alliance with Ana G. M\'endez Universidad Metropolitana and the
  Universities Space Research Association. We have made use of NASA's
  Astrophysics Data System Bibliographic Services, the HyperLeda
  database and the NASA/IPAC Extragalactic Database (NED) which is
  operated by the Jet Propulsion Laboratory, California Institute of
  Technology, under contract with the National Aeronautics and Space
  Administration.}

\bibliographystyle{mn2e.bst}
\bibliography{mybib1,granot}

\clearpage

\end{document}